\documentclass[%
reprint,
superscriptaddress,
showkeys,
amsmath,amssymb,
aip,apl,%
]{revtex4-1}
\pdfoutput=1
\usepackage{placeins}
\usepackage{graphicx}
\usepackage{comment}
\usepackage[caption=false]{subfig}
\usepackage{framed}
\usepackage[normalem]{ulem}
\usepackage[export]{adjustbox}
\usepackage{dcolumn}
\usepackage{bm}
\usepackage{array,ragged2e}
\newcolumntype{P}[1]{>{\Centering\hspace{0pt}}p{#1}}
\usepackage{subfig}
\newsavebox{\measurebox}
\usepackage[utf8]{inputenc}
\usepackage[T1]{fontenc}
\usepackage{mathptmx}
\usepackage[dvipsnames]{xcolor}
\usepackage{textcomp}
\usepackage{amsmath,amssymb,stmaryrd} 
\usepackage{gensymb}

\begin{document}
\preprint{APS/123-QED}
\title{Defect-Moderated Oxidative Etching of MoS$_2$}
\author{Pierce~Maguire}
\affiliation{School of Physics, Trinity College Dublin, Dublin 2, Ireland, D02 PN40}
\affiliation{AMBER Centre, CRANN Institute, Trinity College Dublin, Dublin 2, Ireland, D02 PN40}

\author{Jakub~Jadwiszczak}
\affiliation{School of Physics, Trinity College Dublin, Dublin 2, Ireland, D02 PN40}
\affiliation{AMBER Centre, CRANN Institute, Trinity College Dublin, Dublin 2, Ireland, D02 PN40}

\author{Maria~O'Brien}
\affiliation{AMBER Centre, CRANN Institute, Trinity College Dublin, Dublin 2, Ireland, D02 PN40}
\affiliation{School of Chemistry, Trinity College Dublin, Dublin 2, Ireland, D02 PN40}

\author{Darragh~Keane}
\affiliation{AMBER Centre, CRANN Institute, Trinity College Dublin, Dublin 2, Ireland, D02 PN40}
\affiliation{School of Chemistry, Trinity College Dublin, Dublin 2, Ireland, D02 PN40}
\date{\today}

\author{Georg~S.~Duesberg}
\affiliation{AMBER Centre, CRANN Institute, Trinity College Dublin, Dublin 2, Ireland, D02 PN40}
\affiliation{School of Chemistry, Trinity College Dublin, Dublin 2, Ireland, D02 PN40}
\affiliation{Institute of Physics, EIT 2, Faculty of Electrical Engineering and Information Technology, Universit{\"a}t der Bundeswehr M{\"u}nchen, Werner-Heisenberg-Weg 39, 85577 Neubiberg, Germany}

\author{Niall~McEvoy}
\affiliation{AMBER Centre, CRANN Institute, Trinity College Dublin, Dublin 2, Ireland, D02 PN40}
\affiliation{School of Chemistry, Trinity College Dublin, Dublin 2, Ireland, D02 PN40}

\author{Hongzhou~Zhang}
\email{Hongzhou.Zhang@tcd.ie}
\affiliation{School of Physics, Trinity College Dublin, Dublin 2, Ireland, D02 PN40}
\affiliation{AMBER Centre, CRANN Institute, Trinity College Dublin, Dublin 2, Ireland, D02 PN40}

\begin{abstract}
We report a simple technique for the selective etching of bilayer and monolayer MoS$_2$. In this work, chosen regions of MoS$_2$ were activated for oxygen adsorption and reaction by the application of low doses of He$^+$ at 30 keV in a gas ion microscope. Raman spectroscopy, optical microscopy and scanning electron microscopy were used to characterize both the etched features and the remaining material. It has been found that by using a pre-treatment to introduce defects, MoS$_2$ can be etched very efficiently and with high region specificity by heating in air. 
\end{abstract}
\flushbottom
\maketitle

\thispagestyle{empty}

\section{Introduction}
\begin{table*}
    \centering
    \begin{tabular}{l l l l l l l l l}
     Ref.               & Synthesis & Layer No.   & $T_{min}$        &$T_{ads}$      & Etchant & Characterisation  & Oxide\\ \hline 
     \cite{Wu2013a}     & Mech. Exf.& 1, 2, 4L & $330$\degree{}C  & -             & Air     & AFM $\&$ MFM   & MoO$_3$\\
     \cite{Zhou2013a}   & Mech. Exf.& 1-4L     & $345$\degree{}C  & -             & Air     & AFM $\&$ Raman & None\\
     \cite{Zhou2013a}   & Mech. Exf.& 10L+     & $345$\degree{}C  & -             & Air     & AFM $\&$ Raman & None\\
     \cite{Yamamoto2013}& Mech. Exf.& 1-4L     & $320$\degree{}C  &$200$\degree{}C & Ar/O$_2$& AFM $\&$ Raman & None\\
     \cite{Yamamoto2013}& Mech. Exf.& $>$40nm  & $400$\degree{}C  & -             & Ar/O$_2$& AFM $\&$ Raman & MoO$_3$\\
     \cite{Neupane2016} & CVD       & 2-3L     & $240$\degree{}C  & -             & O$_2$   & N/A           & N/A  \\
     \cite{Rao2017}     & CVD       & Few, $\sim$3L&$300$\degree{}C & -            & Ar/O$_2$& Raman          & None \\
     &
    \end{tabular}
    \caption{Key results from the literature concerning the oxidation of MoS$_2$. This table lists the following details from each study: synthesis method, layer number of starting material, minimum temperature at which the oxidation reaction is reported to occur ($T_{min}$), minimum temperature at which adsorption effects were reported ($T_{ads}$), the oxidative species to which the sample was exposed, the characterisation method used in the search for oxide material and finally the authors' conclusion on whether solid oxide material remained after etching.}
    \vspace{0ex}
    \label{table:oxidationKeyPapers}
\end{table*}

For 2D transition metal dichalcogenides (TMDs) to become prolific in devices, their stability in diverse chemical and physical environments must be understood and highly scalable processing must be available. MoS$_2$ is one such material which exhibits diverse properties in its various forms \cite{Kam1982, Mak2010, Lee2010, Splendiani2010, Liu2013}, allowing potential applications in flexible electronics \cite{Pu2012}, photodetectors \cite{Lopez-Sanchez2013b} and solar cells \cite{Zhang2014}. MoS$_2$ synthesized by chemical vapor deposition (CVD) can be produced in relatively large crystals (several micrometers) with control over the density of defects and layer number \cite{Zhan2012, Lee2012, Song2015}. However, sulfur vacancies are inevitable, even in high quality samples and a native n-type doping is typically noted \cite{Qiu2013,Fivaz1967}.
The oxidation of bulk or powdered MoS$_2$ in air to MoO$_x$ has been well scrutinized because of profound impact on its performance as an industrial lubricant---transforming electronic, chemical, optical, and tribological properties \cite{ballou1953,Ross1955,Lansdown1999}. When heated in the presence of oxygen, the reaction forms MoO$_3$ and SO$_2$ molecules:
\begin{equation}
\mathrm{MoS}_{2(s)} + 3.5\mathrm{O}_{2(g)} \rightarrow \mathrm{MoO}_{3(s/g)} +2\mathrm{SO}_{2(g)}
\label{eq:oxidationequation}
\end{equation}
The temperature at which the reaction occurs is greatly determined by the condition of the material, reportedly ranging from 100\degree{}C for a powder \cite{Ross1955}, to 400\degree{}C \cite{Khare2013} for sputtered MoS$_2$ coatings. The oxidative thinning and/or etching of 2D MoS$_2$ can be achieved by heating in the presence of O$_2$ to $\sim$240\degree{}C or higher \cite{Yamamoto2013, Zhou2013a, Wu2013a, Gan2016, Neupane2016}. Other oxidants have also been used such as the more reactive O$_3$ \cite{yamamoto2013oxidation, Azcatl2014}, XeF$_2$ \cite{Huang2013} and oxygen containing plasmas \cite{Zhu2016, Jadwiszczak2017, Jadwiszczak2019}. Many reports demonstrate a mesh of quasi-equilateral triangular pits in the MoS$_2$ surface after exposure to oxidizing conditions \cite{Wu2013a,Zhou2013a,Yamamoto2013,Gan2016}. These pits are likely bounded by the p-doped zig-zag-Mo edge, with each Mo atom bonded to two O atoms in a wide range of O chemical potentials \cite{Schweiger2002, Jaramillo2007, Zhou2013a, Gan2016}. At temperatures above $\sim$250\degree{}C, the density of etched pits depends on the availability of defect sites about which to nucleate. Pit density does not correlate strongly with the environmental conditions \cite{Yamamoto2013,yamamoto2013oxidation}. Furthermore, kinetic Raman spectroscopy and DFT studies suggest that the reaction energy for defective MoS$_2$ is much lower than the value calculated for pristine material \cite{Rao2017,Kc2015}.
The sublimation temperature for bulk MoO$_3$ is normally about 700\degree{}C but that value can be substantially less for nanoscale MoO$_3$ or MoO$_x$ as a reaction product \cite{Ross1955, Lince2000, Hu2015}. In reports of bulk, thicker films ($\gtrsim$40 nm) and some powdered MoS$_2$ the oxidized molybdenum remains and its relative content can be measured \cite{Ross1955,Lince2000,Yamamoto2013}. Wu \textit{et al}. used magnetic force microscopy and atomic force microscopy to find evidence of MoO$_3$ after heating few-layer MoS$_2$ \cite{Wu2013a}. However, evidence for the presence of oxide material is mixed for thin samples (1-4L) and the precise conditions for oxidative experiments in which the MoO$_x$ reaction product sublimes or endures remain understudied \cite{Yamamoto2013, Zhu2016}. It seems that MoO$_3$ sublimation dominates for some kinetic and material conditions (especially thin samples). Heating experiments and results from papers discussed here are summarized in table \ref{table:oxidationKeyPapers}.


\begin{figure*}
\centering
\captionsetup[subfigure]{labelformat=empty}
\subfloat[][]{\label{oxidationschematic}\includegraphics[width=6.00in]{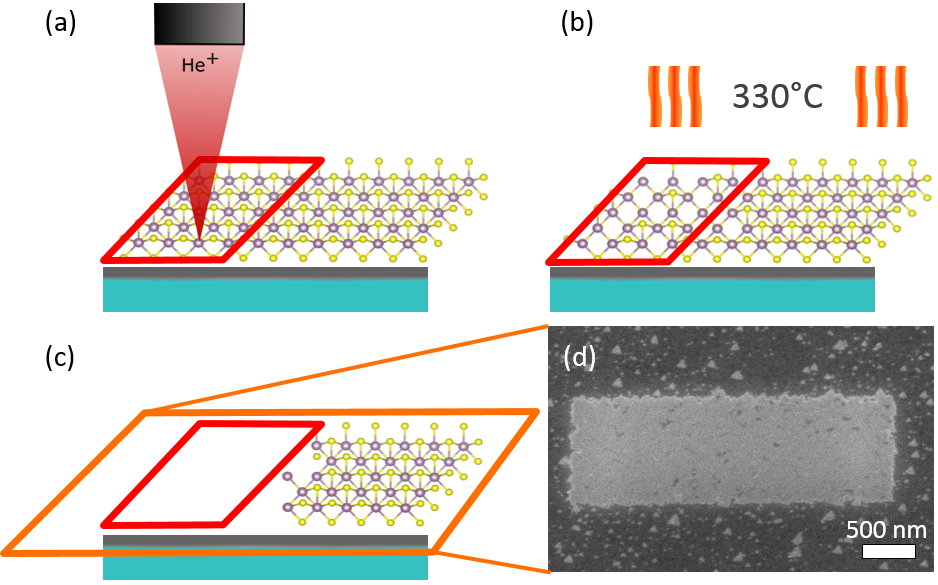}}
\caption{Outline of a typical defect-moderated oxidative etching experiment.
(a) illustrates the 30 keV He$^+$ beam incident on a rectangular region of a CVD grown film of MoS$_2$.
(b) shows defective MoS$_2$ undergoing chemical etching in air at 330\degree{}C. 
(c) shows the same region, now etched.
(d) is a SEM micrograph of one such etched rectangle.}
\label{fig:oxidationschematic}
\end{figure*}

Raman spectroscopy has been employed extensively in the characterization of MoS$_2$ in various forms such as bulk \cite{Wieting1971, Chen1974, Windom2011, Livneh2015}, powder \cite{Stacy1985}, nanoparticles \cite{Frey1999, Rice2013} and few-layer/monolayer \cite{Lee2010, Chakraborty2012, Li2012, Li2012a, Zhao2013}. The high-energy optical modes that are Raman active in monolayer/bilayer MoS$_2$ are the following: the $E'$/$E_{g}$ peak at $\sim$385 cm$^{-1}$ which arises from the intralayer, in-plane motion of Mo and S atoms with respect to each other and the $A'_{1}$/$A_{1g}$ peak at $\sim$405 cm$^{-1}$ which arises from the intralayer, out-of-plane motion of S atoms \cite{Lee2010, Li2012}. The frequency and width of the $A'_{1}$/$A_{1g}$ peak has been reported to be sensitive to electrostatic doping while the position, splitting and width of $E'$/$E_{g}$ peak is more sensitive to strain \cite{Chakraborty2012,Rice2013,McCreary2016}. It has been shown using symmetry arguments, DFT calculations and experiments that an upshift in the frequency and decrease in the linewidth of the Raman $A'_1$/$A'_{1g}$ modes represent a strong electron interaction for that optical phonon. Electron density is significantly diminished by oxygen treatment above $200$\degree{}C causing these effects on the $A'_1$/$A'_{1g}$ mode \cite{Chakraborty2012, Shi2013, Wu2013a, Yamamoto2013}. Since etching/thinning of 2D MoS$_2$ has not been found at temperatures of $\sim$200\degree{}C, narrowing and upshifting of the $A'_1$/$A'_{1g}$ modes at temperatures below $200$\degree{}C are caused by increased adsorption rather than oxidation. 
Neupane \textit{et al.} performed experiments with dry O$_2$ gas on CVD MoS$_2$ and demonstrated in Raman and photoluminescence spectroscopies that annealing in H$_2$ caused a reversal of the effects of adsorbed oxygen on the doping state of the remaining material \cite{Neupane2016}. The adsorption energy of molecular oxygen on the monolayer MoS$_2$ surface is reduced by approximately half in the presence of a surface sulfur vacancy defect. Therefore, controlling defect density is critical in controlling the adsorption of oxygen and hence oxidation \cite{Kc2015}. 

The precise defect-engineering of 2D materials has been demonstrated using highly spatially resolved ion irradiation in a helium ion microscope (HIM) \cite{Fox2015,Klein2017,Maguire2018,Maguire2019, Iberi2016, Stanford2016, Nanda2017a, Zhou2016a}. In this paper, we use such methods to moderate adsorption and the oxidative reaction of MoS$_2$ in air, demonstrating a high degree of spatial control over the oxidation reaction, as illustrated schematically in figure \ref{fig:oxidationschematic}. This work facilitates high throughput patterning for preferential oxygen adsorption and oxidation. Our Raman spectroscopy results indicate that the selective adsorption of oxygen at defect sites could also be used to create a localized p-type doping environment. In this paper, unprecedented spatial control of the oxidation reaction by pre-treating with an ion beam is demonstrated, and several aspects of the reaction are clarified e.g. the presence of MoO$_x$, the effects of temperature, and the influence of ion dose.

\section{Methods}
MoS$_2$ was prepared using a previously described CVD technique \cite{OBrien2014}. The MoS$_2$ thickness was checked using the peak separation of the $A_1'$/$A_{1g}'$ and $E'$/$E^1_{2g}$ peaks in Raman spectroscopy \cite{Lee2010}. The \textit{Zeiss ORION NanoFab} microscope was used to irradiate MoS$_2$ with He$^+$ at an energy of 30 keV and an angle of incidence of 0$^{\circ}$. Various arrays were irradiated as detailed in the results and discussion. The dwell time, number of scans and beam current were varied to ensure that the desired dose was delivered. Beam currents used were between 1 and 4.2 pA. 

The heating experiments were performed by loading samples on a glass slide into the middle of an \textit{MTI Multi-Position GSL-1100X-NT-UL-LD} quartz tube furnace. The furnace was sealed, containing only air at atmospheric pressure without any flow. The temperature was raised to the desired etching temperature at a rate in the range of 10 $\pm$ 2\degree{}C min$^{-1}$. The sample was heated to its maximum temperature (e.g. 330\degree{}C) at which it was held for the desired time. After being allowed to cool to room temperature naturally, the sample was imaged using optical microscopy. A schematic of a representative ion-moderated etching experiment is shown in figure \ref{fig:oxidationschematic}.

Ex-situ Raman spectroscopy was carried out using a \textit{WITec Alpha 300R} system (532 nm laser) with a 100$\times$ objective (NA=0.95) (spot size $\sim$0.3 $\mathrm{\mu m}$) and a 1800 lines/mm grating. Raman maps were generated by taking four spectra per $\mathrm{\mu m}$ in both the x and y directions over large areas \cite{Delhaye1975}. Each acquisition was for 0.05 s. The laser power was $\sim$1 mW or less to minimize damage to the samples. The spectra from a desired region were acquired by averaging. Lorentzian distributions were fitted to the Raman peaks as indicated in Figure 3.

Ex-situ scanning electron microscopy (SEM) was performed with a field emission SEM (\textit{Zeiss Supra} fitted with a \textit{Gemini} column, \textit{Zeiss Microscopy GmbH}, Jena, Germany). The microscope was operated at a beam energy of 10 keV.
\section{Results}
\begin{figure*}
\captionsetup[subfigure]{labelformat=empty}
\centering
\subfloat[]{\label{oxdosedependancy1}\includegraphics[height=2.20in]{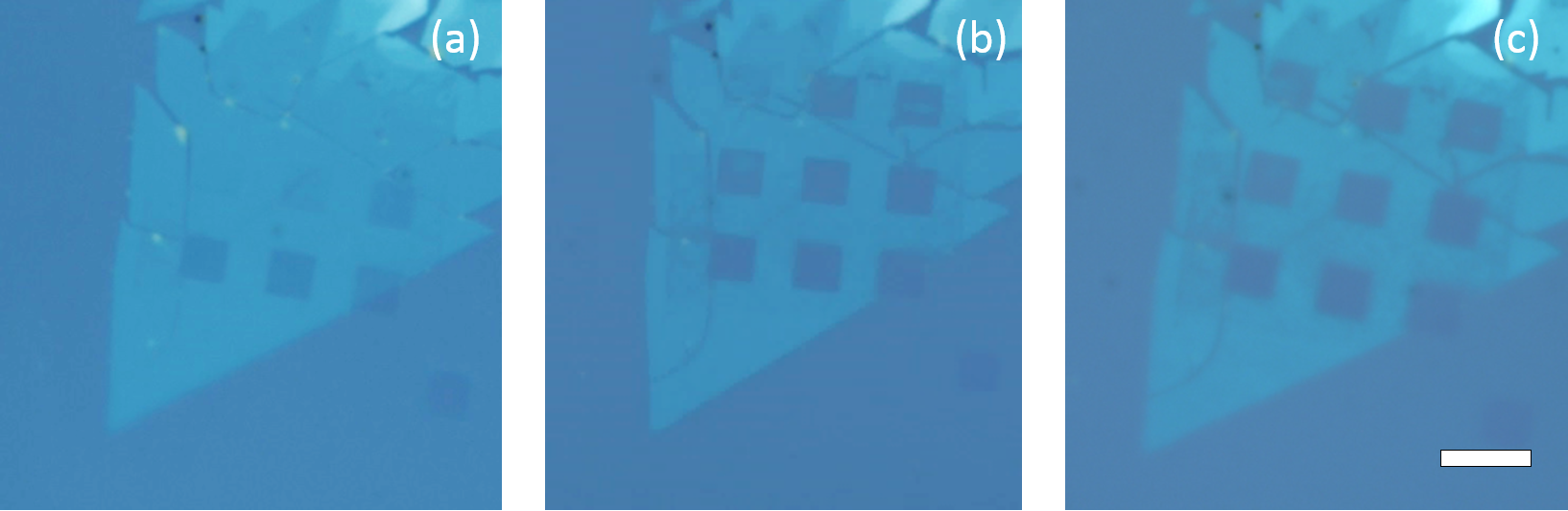}}\qquad
\caption{Optical images showing the oxidative etching of bilayer CVD MoS$_2$ irradiated with a variety of ion doses. (a) includes nine $5 \mathrm{\mu m}\times5$ $\mathrm{\mu m}$ irradiated squares in the MoS$_2$ flake. The dose of each square increases from left to right starting in the top row. The doses are $1\times 10^{13}$,
$5\times 10^{13}$,
$1\times 10^{14}$,
$5\times 10^{14}$,
$1\times 10^{15}$,
$5\times 10^{15}$,
$1\times 10^{16}$,
$5\times 10^{16}$ and 
$1\times 10^{17}$ He$^+$ cm$^{-2}$. Only the four highest doses are clearly distinguishable before heating. 
(b) The same region after heating for 50 minutes at 330\degree{}C. (c) The same region after heating for 50 minutes at 330\degree{}C for a second time. The scale bar is 10 $\mathrm{\mu m}$.}
\label{fig:oxdosedependancy}
\end{figure*}

Figure \ref{fig:oxdosedependancy}(a) shows optical microscope images of an array of $5 \mathrm{\mu m}\times5$ $\mathrm{\mu m}$ squares in a bilayer region of CVD MoS$_2$ which were irradiated with 30 keV He$^+$. The irradiation doses range from $5\times 10^{13}$ He$^+$ cm$^{-2}$ (which is not visible, in the top left) to $1\times 10^{17}$ He$^+$ cm$^{-2}$ (notably discolored, in the bottom right row). There is another square of irradiated substrate which is barely visible to the bottom right. Figure \ref{fig:oxdosedependancy}(b) shows the same region after heating for 50 mins at 330\degree{}C. All of the squares on MoS$_2$ are now much more clearly visible and most (exluding the lowest doses) appear much more like the irradiated substrate, suggesting they have been etched. The minimum dose which resulted in etching of the MoS$_2$ appears to be $1 \times10^{14}$ He$^+$ cm$^{-2}$ although some regions with doses up to $5\times10^{14}$ He$^+$ cm$^{-2}$ were not fully etched. Doses above this were always sufficient to remove all material in a region after heat treatment. In addition, when the heating step was repeated for a further 50 minutes under the same conditions, even the lowest dose ($5\times10^{13}$ He$^+$ cm$^{-2})$ region was almost completely etched, as shown in figure \ref{fig:oxdosedependancy}(c). 

\begin{figure*}
\centering
\subfloat[][Not irradiated]{\label{sub:0MoS2}\includegraphics[width=3.00in]{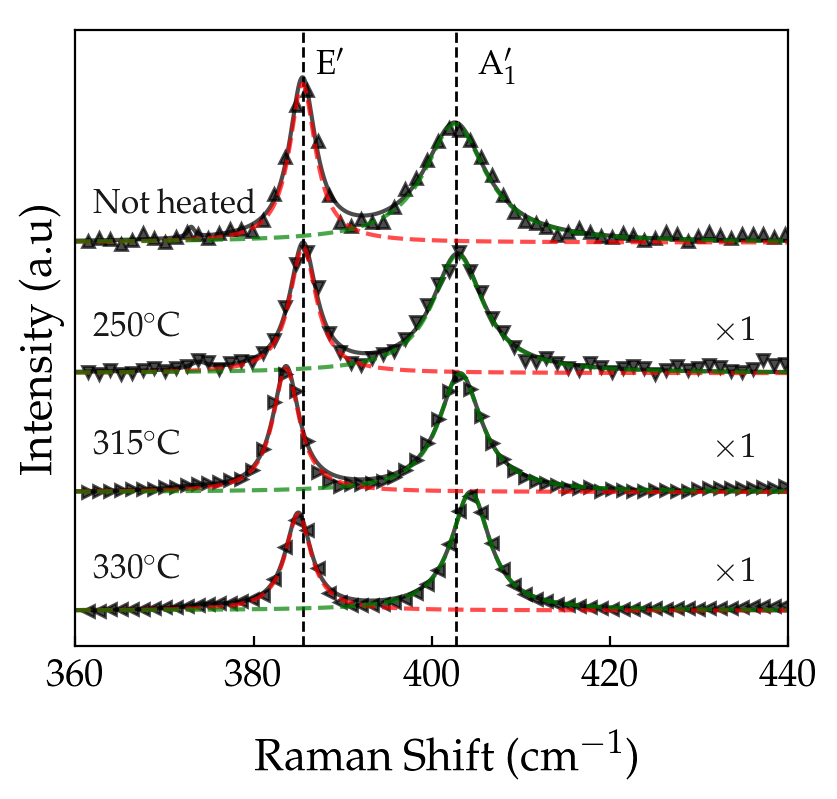}}
\subfloat[][Irradiated with $5\times10^{14}$ He$^+$ cm$^{-2}$]{\label{sub:5E14MoS2}\includegraphics[width=3.00in]{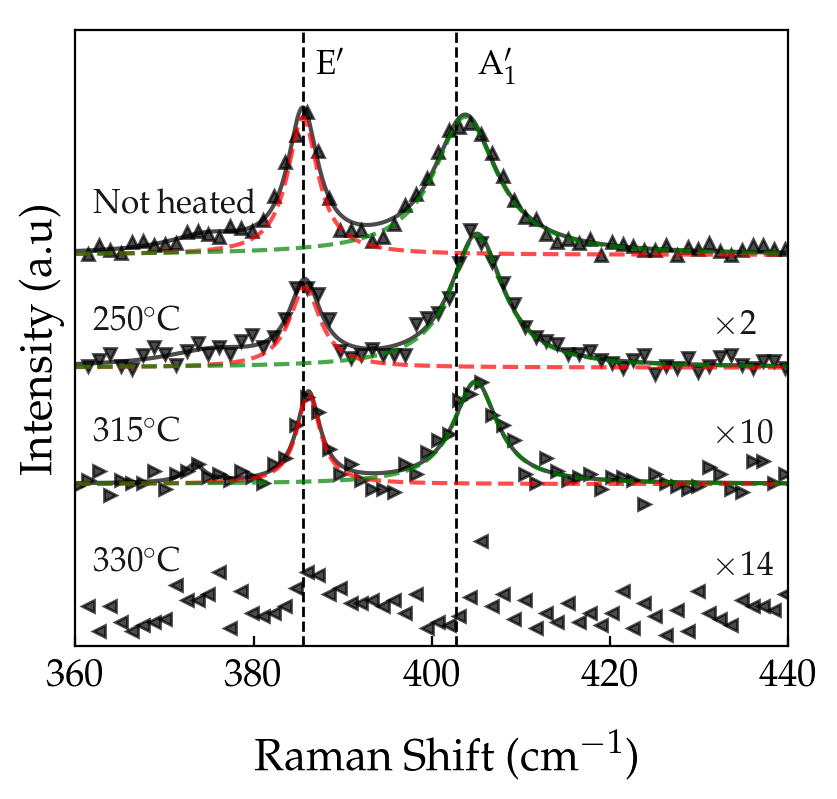}}\qquad
\caption{Raman spectroscopy of monolayer CVD MoS$_2$ with irradiation and heating. (a) and (b) show sets of Raman spectra for non-irradiated MoS$_2$ and MoS$_2$ irradiated with a dose of $5\times10^{14}$ He$^+$ cm$^{-2}$ respectively. The offset spectra are from samples which had been heated to the indicated temperature for 50 mins in air. The triangles represent individual data points, the black line is the fit to all of the data, and the red and green lines are fits to the $E'$ and $A_1'$ peaks respectively. The $E'$ and $A_1'$ peaks are labelled, and their original (non-irradiated, non-heated) positions are indicated by the dashed black lines on both figures. Each spectra is normalized to the maximum of the non-irradiated $A_1'$ peak. The irradiated spectra are multiplied by the factor indicated indicated on the right for visibility.}
\label{figure:mechanismraman}
\end{figure*}

A new array of rectangles was irradiated in monolayer regions of CVD MoS$_2$ for investigation with Raman spectroscopy. The irradiated patterns are overlaid on a low dose HIM image which is presented in Figure S1 along with Raman maps. Figure \ref{figure:mechanismraman} shows two selected sets of MoS$_2$ Raman spectra, both non-irradiated and irradiated (5 $\times$ 10$^{14}$ He$^+$ cm$^{-2}$), before and after heating to different maximum temperatures. For the non-irradiated spectra in figure 3(a), a slight increase in peak separation and a narrowing of the $A_1'$ peak are observed as the maximum temperature increases. For the irradiated spectra in figure 3(b), the results are more dramatic. As expected, the peak separation is increased even without heating, and it is further increased by heating in air. The Raman intensity is observed to decline sharply with heating and there are no clearly identifiable peaks after heating to 330\degree{}C. It should be noted that no residual oxides of molybdenum could be detected in the etched regions, suggesting the complete sublimation of MoO$_x$ (see Figure S2) \cite{Wachs2010,Diskus2012,Yang2012b,Lafuente2015}. 

\begin{figure*}
\centering
\captionsetup[subfigure]{labelformat=empty}
\subfloat[][]{\label{sub:semworkfunctionRT}\includegraphics[height=1.90in]{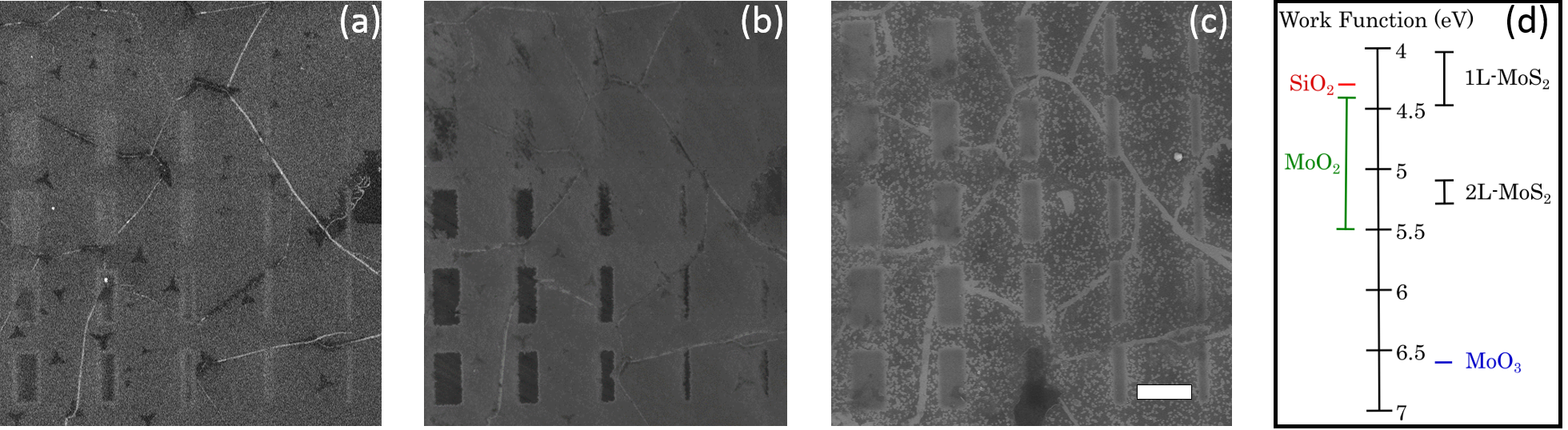}}\hspace{0.1in}
\caption{SEM imaging of monolayer MoS$_2$ in a cumulative heating experiment. (a) to (c) is a selection of SEM micrographs at various steps of the cumulative heating process. (a) is after irradiation and before any heating. (b) is after heating to the 305\degree{}C step. (c) is after the 330\degree{}C heating step. The scale bar is 2 $\mathrm{\mu m}$ for all three images. (d) shows the work functions of pertinent materials.}
\label{figure:workfunction}
\end{figure*}

In order to further clarify the effects at different temperatures, another sample of monolayer CVD MoS$_2$ was prepared and an array of rectangles of various widths and doses was irradiated in it. The sample was put through a series of heating steps lasting 10 mins at a maximum temperature which was iteratively increased from 300\degree{}C to 335\degree{}C in 5\degree{}C increments. Between each heating step the sample was imaged in a field emission SEM using a through-the-lens (TTL) or "inLens" detector, chosen for its particular sensitivity to the work function of the specimen. A selection of the SEM micrographs are shown in figure \ref{figure:workfunction}. They demonstrate the progression from irradiated MoS$_2$ through to the complete etching of irradiated material after heating. Figure \ref{figure:workfunction}(a) shows low dose irradiated regions which are lighter than surrounding areas and higher dose regions which are slightly darker. After the 305\degree{}C step, figure \ref{figure:workfunction}(b) shows the same regions where those that received lower doses are still relatively unchanged and those that received higher doses are now very dark. Figure \ref{figure:workfunction}(c) shows the same regions once again after the 330\degree{}C step where all irradiated regions now appear to have been etched.
Using an inLens detector, low work function materials typically appear bright and high work function materials appear dark. Changes in the signal intensity of a material in SEM images acquired in these conditions can be used to qualitatively infer changes in the material work function. The work function of monolayer MoS$_2$ ($\sim$4.0 eV) has been reported to increase as O$_2$ becomes adsorbed to its surface \cite{Lee2016a}. The work functions of MoO$_2$ and MoO$_3$ are $\sim$4.4-5.5 eV and 6.6 eV respectively \cite{Bernhard1999,Moosburger-Will2009,Guo2014}.The range of work functions of possible materials in this system is represented in figure \ref{figure:workfunction}(d). The work function is expected to increase after exposure to oxygen, first because of adsorption (to $\sim$4.5 eV) \cite{Lee2016a}, and possibly again when the material is oxidized to MoO$_x$ if it remains in place ($\sim$4.4-6.6 eV). The increased work function causes a darkening of the sample evident in figure \ref{figure:workfunction}(b), although the distinction between adsorption and oxidation is not initially clear here, as will be discussed further below. Should a region be observed to lighten after a heating step, it must be due to the oxidation reaction coupled with evaporation of the MoO$_x$ species, revealing the SiO$_2$ substrate underneath which has a work function more similar to the starting material (4.3 eV) \cite{Yan2012a}. 

\begin{figure*}
\captionsetup[subfigure]{labelformat=empty}
\centering
\subfloat[]{\label{sub:2e14}\includegraphics[height=4.9in]{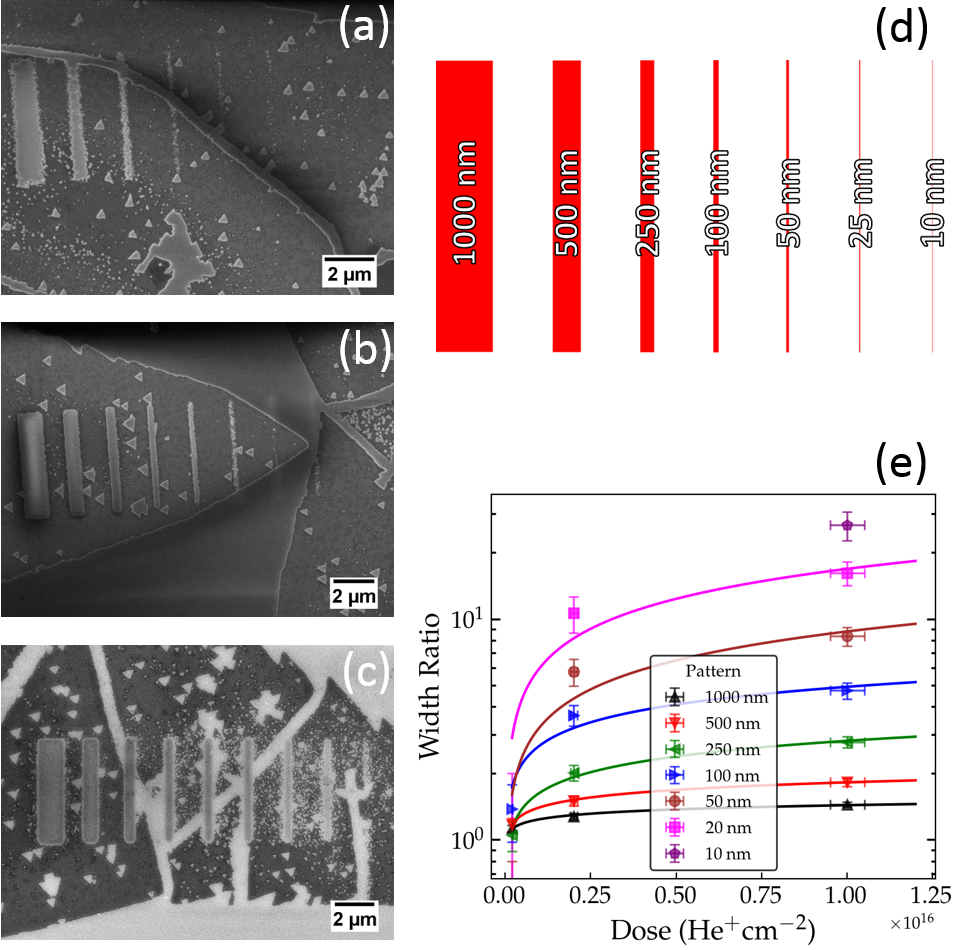}}\quad
\caption{Width dependence of etched features. (a),(b) and (c) show SEM images of arrays irradiated with 2 $\times10^{14}$ He$^+$ cm$^{-2}$,  2 $\times10^{15}$ He$^+$ cm$^{-2}$ and 1 $\times10^{16}$ He$^+$ cm$^{-2}$ respectively. (d) shows a schematic diagram of the array pattern with decreasing widths from left to right. (e) shows a graph of the etched width as measured using SEM against irradiation dose for the different irradiation widths. The higher doses have created larger etched features. The fitted curve for each irradiated width is from equation \ref{eq:widths}.} 
\label{fig:widths}
\end{figure*}

To investigate the effect of ion dose on the size of etched features, arrays of rectangles with descending widths were irradiated in monolayer MoS$_2$ for different doses and etched at 330\degree{}C for 50 mins. These arrays are shown in figure \ref{fig:widths}(a),(b) and (c) for 2 $\times$ $10^{14}$ He$^+$ cm$^{-2}$,  2$\times$ $10^{15}$ He$^+$ cm$^{-2}$ and 1$\times$ $10^{16}$ He$^+$ cm$^{-2}$ respectively. The rectangle widths are indicated schematically in figure \ref{fig:widths}(d), varying from left to right as follows: 1000, 500, 250, 100, 50, 20, 10, 5, 1 nm. 
Even the smallest width can be seen for the highest dose but only rectangles with irradiated widths of $\sim$250 nm or above are fully etched for the lowest dose. Narrower rectangles for the lowest dose are observed to be partially etched, being not contiguous. Other etching is observed to occur along previously hidden grain boundaries in figure \ref{fig:widths}(a) and (c). 
Figure \ref{fig:widths}(e) shows a graph of the width ratio ($W_R$), defined as the SEM-measured width normalised to its designed width, against the delivered dose ($D$) for rectangles of different widths. Larger doses are required to ensure that the smallest patterned features are fully etched. Once above the threshold for etching, the effect of increasing the dose is to broaden the etched feature. The relationship between $W_R$ and $D$ is of the form: 
\begin{equation}
W_R(D) = \mathrm{a}D^{\mathrm{b}}
\label{eq:widths}
\end{equation}
where $\mathrm{a}$ and $\mathrm{b}$ are fitting parameters. 
From the 20 nm to the 1000 nm designed width rectangles, $a$ increases exponentially from $2\times10^{-5}$ to 136. 
At the same time,
$\mathrm{b}$ decreases exponentially from 0.45 to 0.064.

\section{Discussion}
The ion dose demonstrated to induce etching ($\sim10^{14}$cm$^{-2}$) corresponds to an ion-induced defect density of about one in a thousand atoms. It is below the dose normally required to remove material by direct ion milling by $\sim$4 orders of magnitude. It was also less than the dose required for amorphisation or detectable change in stoichiometry by 2 to 4 orders of magnitude \cite{Fox2015}.

\begin{figure*}
\centering
\captionsetup[subfigure]{labelformat=empty}
\subfloat[Part 1][]{\label{sub:SEMoxidationheatmap}\includegraphics[width=0.950\textwidth,valign=t]{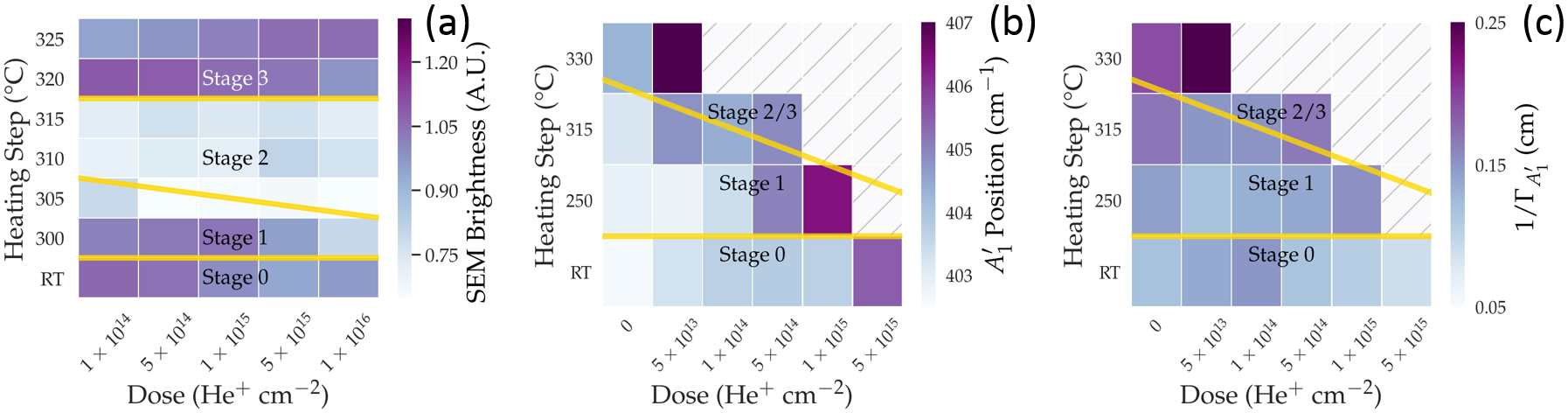}}
\caption{Heat maps of SEM brightness and $A'_1$ peak position and width for monolayer MoS$_2$.
(a) shows SEM brightness for irradiated regions--- normalised to the brightness of nearby non-irradiated material---as a function of their cumulative heating step and irradiation dose.
(b) and (c) show the position and inverse width of the $A'_1$ peak respectively as a function of the heating temperature and ion dose.}
\label{figure:oxidationheatmaps}
\end{figure*}

Figure \ref{figure:oxidationheatmaps} shows several heat maps, representing a summary of salient data from the Raman and SEM experiments. We can use them to label several discrete stages separated by their temperature range. 
\textbf{Stage 0} is the initial state of the irradiated MoS$_2$, before the application of any heating step. When characterized, a dose-dependent increase in the frequency and width of the A$'_1$ peak along with an apparent increased work function are observed \cite{Mignuzzi2015, Maguire2018, Klein2017, Maguire2019}. From the literature, these observations are consistent with the material being increasingly defective. 

\begin{table}
\centering
\begin{tabular}{|l|l|l|l|l|l|l|l|l|}
\hline
\# & Temp (\degree{}C) & $\omega_{A'_1}$    & $\Gamma_{A'_1}$   &  $W$            & Description               \\ \hline
0    & RT           & All $\uparrow$    & H.D. $\uparrow$ & H.D. $\uparrow$ & Increased Defects \\
1    & $<305$  & All $\uparrow$    & H.D. $\downarrow$ & H.D. $\uparrow$ & O$_2$ Adsorption at Defect Sites\\
2   & $305-320$& All $\uparrow$    & All $\downarrow$ & All $\uparrow$ & O$_2$ Adsorption to Saturation\\
3 & $>320$\   & All --        	& All --      & All $\downarrow$& Oxidation\\
\hline
\end{tabular}
\caption{Temperature dependent stages of heating monolayer MoS$_2$ in air. 'All' means all doses, H.D. means high doses, $\uparrow$ and $\downarrow$ indicate increases and decreases respectively, $\omega_{A'_1}$ is the position of the $A'_1$ peak, $\Gamma_{A'_1}$ is the width of the $A'_1$ peak, $W$ is the work function.}
\label{table:oxidationstages}
\end{table}

\textbf{Stage 1} is applicable after heating to temperatures $<$305\degree{}C. Unlike in \textbf{stage 0}, the position and width of the $A'_1$ peak no longer increase together. While the frequency continues to increase, the $A'_1$ peak is now observed to narrow and SEM brightness is slightly decreased. Oxygen adsorption has an electron withdrawing effect on adjacent MoS$_2$. Therefore, these are clear indications of increased oxygen content and resulting p-type doping \cite{Chakraborty2012,Wu2013a,Yamamoto2013}. In this range, oxidative etching is rarely noted in the literature and only for times much longer than those used in this work. The increased defect density caused by the ion irradiation has lowered the barrier for oxygen adsorption \cite{Fox2015,Kc2015}. Since these effects show such a strong dose relationship in \textbf{stage 1}, it seems that this stage is dominated by activity at the induced defect sites. 

The regime between the temperatures of $305$\degree{}C and $\sim320$\degree{}C is labelled as \textbf{stage 2}. Here, even the $A'_1$ peak of non-irradiated and low dose MoS$_2$ is now strongly affected. In addition, the apparent change in work function is now much more intense. This is attributed to a highly increased oxygen adsorption rate which is no longer dependent on defects as was the case in \textbf{stage 1}. The first etched pits in our non-irradiated material are observed at the top of this range suggesting that oxidation is beginning to occur significantly.

\textbf{Stage III} occurs at 320-325\degree{}C. By this stage, it is observed in the SEM images that brightness is now similar across all irradiated areas regardless of dose. 
This stage is therefore attributed to the complete oxidation and sublimation of Mo species leaving behind only the SiO$_2$ substrate. The key characteristics of these stages are presented in table \ref{table:oxidationstages}.

At the lower temperatures, the effect of ion dose is strong. However, at temperatures above $\sim$320\degree{}C, the distinction between irradiated and non-irradiated material begins to matter less and even non-irradiated MoS$_2$ begins to etch. This can be seen in the SEM image in figure \ref{figure:workfunction}(c) and is also clear from Raman spectra results which show even the lowest doses having been fully etched after high temperature treatment. Here, curved lines of material near grain boundaries and many randomly positioned pits at native defect sites are observed to have been etched. 

\section{Conclusion}
In this work, ion irradiation has been successfully established as a high throughput tool for moderating the oxygen adsorption and oxidation of 2D MoS$_2$. Previous studies have used oxidative etching for the doping and etching of MoS$_2$ but this is the first report of doing so while asserting spatial control, leaving other non-irradiated regions relatively untouched. Region-specific, low dose and high throughput irradiation with He$^+$ at 30 keV allowed thermal etching to be preferentially stimulated in discretionary regions. At heating temperatures comparable to the literature, doses as low as 1 $\times10^{14}$ He$^+$ cm$^{-2}$ showed a profound effect on oxidative etching. This method also avoids the use of resists which are a leading source of contamination in 2D material based devices. A breakdown of effects at different temperatures using Raman spectroscopy and SEM was also provided, developing understanding of the stability of defective MoS$_2$ in potentially oxidative environments. 

\section*{Additional information}

\textbf{Supplementary Information} accompanies this paper at **. It includes Raman maps and Raman spectra in relation to the absence of oxides.

\section*{Acknowledgements}
The authors thank the staff at the Advanced Microscopy Laboratory (AML), CRANN, Trinity College Dublin. We acknowledge support from the following grants: Science Foundation Ireland [grant numbers: 12/RC/2278, 15/SIRG/3329, 11/PI/1105, 07/SK/I1220a, 15/IA/3131 and 08/CE/I1432] and Irish Research Council [grant number: GOIPG/2014/972]. Figure 1 was prepared with use of the \textit{Vesta} software package \cite{vesta2011}.

\section*{Bibliography}

\bibliography{sample}



\end{document}